\documentclass{sig-alternate-05-2015}
\usepackage{graphicx}

\usepackage{amsmath}
\usepackage{amsfonts}
\usepackage{amssymb}
\usepackage{color}
\usepackage{floatrow, makecell}
\usepackage{subfigure}
\newtheorem{theorem}{Theorem}

\newcommand{\cale}{\mathcal{E}}
\newcommand{\calf}{\mathcal{F}}
\newcommand{\calg}{\mathcal{G}}

\newcommand{\cali}{\mathcal{I}}

\newcommand{\caln}{\mathcal{N}}

\newcommand{\calq}{\mathcal{Q}}

\newcommand{\calz}{\mathcal{Z}}

\newcounter{protocolcount}

\begin{document}

\title{Joining Local Knowledge to Communicate Reliably (Extended Abstract) \titlenote{The present article includes results from~\cite{DISTPPS17,PPS17}, which appeared in the Distributed Computing journal and FCT 2017 respectively.} }

%\author{Aris Pagourtzis$^1$ \and Giorgos Panagiotakos$^2$ \and Dimitris Sakavalas$^1$}
%
%\date{}

%\title{Reliable Message Transmission under Partial Knowledge and General Adversaries}

\numberofauthors{3}

\author{
% 1st. author
\alignauthor
Aris Pagourtzis\\
       \affaddr{School  of Electrical and Computer Engineering }\\
       \affaddr{ Nat. Tech. Univ. of Athens}\\
%       \affaddr{15780 Athens, Greece}\\
       \email{pagour@cs.ntua.gr}
% 2nd. author
\alignauthor Giorgos Panagiotakos\\
       \affaddr{School of Informatics}\\
       \affaddr{University of Edinburgh}\\
%       \affaddr{ 15784 Athens, Greece}\\
       \email{giorgos.pan@ed.ac.uk}
% 3rd. author
\alignauthor
Dimitris Sakavalas\\
       \affaddr{School  of Electrical and Computer Engineering }\\
       \affaddr{ Nat. Tech. Univ. of Athens}\\
%       \affaddr{15780 Athens, Greece}\\
       \email{sakaval@corelab.ntua.gr}
}

\date{}

\maketitle

\begin{abstract}
A fundamental primitive in distributed computing is Reliable Message Transmission (RMT), which refers to the task of correctly sending a message from a party (or player) to another, in a network where some intermediate relays might be controlled by an adversary. We address the problem under the realistic assumption that the topological knowledge of players is restricted to a certain subgraph and specifically study the role of local information exchange in the feasibility of RMT. We employ the \emph{General Adversary} model of Hirt and Maurer and the recently introduced \emph{Partial Knowledge Model} which subsume all known models for the adversary and local knowledge respectively.  Tight feasibility conditions, naturally involving the network topology, the adversary and the local knowledge of players, are presented.

%We present tight feasibility conditions for RMT which naturally involve the network topology, the possible corruption sets and the topology knowledge of the players.  We address the problem under the \emph{General Adversary} model of Hirt and Maurer (1997) and the  recently introduced \emph{Partial Knowledge Model}. The combination 
%
%
%The problem is addressed under the \emph{general adversary} model of Hirt and Maurer (1997) which subsumes  all known models such as global or local threshold adversaries. We employ the recently introduced \emph{Partial Knowledge Model}, which captures any case of initial players' topology knowledge. Combining these two general models we obtain tight solvability conditions for any adversary  distribution and any knowledge level that has appeared in the literature so far.
%
%We address the problem of \emph{Reliable Message Transmission} (RMT), in the general adversary model of Hirt and Maurer (1997), which subsumes 
%Our main contribution is the determination of a necessary and sufficient condition for achieving RMT in the partial knowledge model with a general adversary. We propose the RMT-Partial Knowledge Algorithm (RMT-PKA), which solves RMT whenever this is possible, therefore it is a \emph{unique} algorithm, as defined in~\cite{PP05}. To the best of our knowledge, this is the first unique protocol for RMT against general adversaries in the partial knowledge model. 

\end{abstract}

\section{Introduction}

Ensuring seamless communication between two indirectly connected entities of a network is fundamental for achieving more complex cooperative tasks. Since vulnerability of modern networks is increased  along with their size, there is a strong interest in achieving consistent communication in unreliable environments. This task is captured by the \emph{Reliable Message Transmission} problem (RMT), in which the goal is the correct delivery of message $x_S$ from a sender $S$ to a receiver $R$  even if some of the intermediate nodes-players are corrupted and do not relay the message as agreed. In this work, we study RMT in the presence of a \emph{Byzantine adversary} which can make some players deviate from the protocol arbitrarily, by  blocking, rerouting, or even altering a message that they should normally relay intact to specific nodes.

%may control several nodes. The adversary is able to make the corrupted nodes deviate from the protocol 
%arbitrarily by blocking, rerouting, or even altering a message that they should normally relay intact
%to specific nodes. An adversary with this behavior is referred to as \emph{Byzantine adversary}.
% \cite{PP05}, $t$-locally bounded Adversary model, honest/corrupted dealer, necessary and 
% sufficient condition, approximation

%Nontrivial distributed computing tasks in unreliable environments require a fundamental primitive called \emph{Reliable Message Transmission} (RMT), which refers to the ability to send a message from a node to another despite the presence of corruptions. 

%Achieving reliable communication in unreliable networks is fundamental in distributed computing. Often, certain parties are only indirectly connected, and need to use 

The feasibility of RMT naturally depends on the structure of the network and the corruption capability of the adversary. Thus, the literature has been focused on the deduction of topological conditions which involve the adversary model and prove necessary or sufficient for RMT to be achieved. We stress that another important parameter affecting distributed tasks in contemporary networks is the level of players' knowledge  over the topology and the adversary. To this end, we employ the recently introduced \emph{Partial Knowledge Model}~\cite{DISTPPS17}, which assumes that each player only has knowledge over some arbitrary subgraph of the network.  It encompasses all previous topology knowledge models including the extensively studied full knowledge~\cite{Dol82, KGSR02} and \textsl{ad hoc}~\cite{PP05, TV13} models (in the latter,  a node is only aware of its immediate neighborhood). The motivation for partial knowledge considerations comes from large scale networks, where global estimation of system properties may be hard to obtain, and social networks, where proximity is often correlated with an increased amount of available information. In this context, we specifically study the role of local information exchange in the feasibility of RMT. Regarding the adversary model we address the problem under the \emph{General Adversary model}~\cite{HM97} which subsumes all known models such as global~\cite{DDWY93} or local~\cite{Koo04, TVB15} threshold adversaries and intuitively captures coalitions of adversarial players. The combination of these two quite general models forms the most general setting encountered within the synchronous deterministic model.

We assume a synchronous network represented by a graph $G=(V,E)$ consisting of the player set $V$ (also denoted as $V(G)$) and edge set $E$ which represents undirected authenticated
%\footnote{As usual in the byzantine faults literature, the existence of authenticated channel $(u,v)$, guarantees that 
%once a message is sent from node $u$ to node $v$, the message will be delivered intact to the receiver $v$ and the receiver will be aware of the identity of the sender $u$.
%% i.e. 
%%no tampering of the message or identity spoofing can be performed by the adversary.
%}
 channels between players. 
% The set of neighbors of a player $v$ is denoted with $\caln(v)$.  
%In our study we make use of node-cuts (separators) which disconnect the sender from the receiver and will simply use the term \emph{cut} to denote such a separator. 
%The problem definition follows.
The neighborhood of a node $v$ will be denoted as $\caln(v)$.
We will denote the sender and the receiver nodes as $S, R\in V$ respectively and say that a distributed protocol \emph{achieves RMT}  if by the end of the protocol the receiver $R$ has \emph{decided on} $x_S$, i.e.\ if it has been able to output  the value $x_S$ originally sent by the sender.\vspace{3pt}

\noindent\textbf{The Adversary Model~\cite{HM97}. } In the General Adversary model, the possible corruption sets are defined by the  \emph{adversary structure} $\calz$, which is, a monotone family of subsets of $V$, 
i.e.\ $\calz \subseteq 2^V$, where  all subsets of a set $Z$ are in $\calz$ if $Z \in \calz$. In this work we obtain our results with respect to  a General Byzantine adversary. \vspace{3pt}
%In this work we obtain our results w.r.t. a general byzantine adversary, i.e.,\ a general adversary which can make all the corrupted players deviate arbitrarily from the given protocol.\smallskip

\noindent\textbf{The Partial Knowledge Model~\cite{DISTPPS17}. } 
Each player $v$ only has knowledge of the topology of a certain subgraph $G_v$ of $G$ which includes $v$. Namely considering the family $\calg$ of subgraphs of $G$ we use the \emph{view function} $\gamma: V\rightarrow \calg$, where $\gamma(v)$ represents the subgraph of $G$ over which player $v$ has knowledge of the topology. We extend the domain of $\gamma$ by allowing a set $S \subseteq V(G)$ as an argument. The value $\gamma(S)$ will correspond to the \emph{joint view} of nodes in $S$. More specifically, if $\gamma(v)=G_v=(V_v,E_v)$ then $\gamma(S)=G_S=(\bigcup_{v\in S} V_v,\bigcup_{v\in S}E_v)$. The extensively studied \textsl{ad hoc} model (cf.~\cite{PP05})  can be seen as a special case of the Partial Knowledge Model, where $\forall v\in V, \ \gamma(v)=\caln(v)$.  

To combine these two models, we first define the \emph{restriction} of a structure $\calz$ to  a node set $A$ as  $\calz^A = \{ Z \cap A \mid  Z \in \calz \}$. We assume that given the actual adversary structure $\calz$ each player $v$ is aware of its \emph{local adversary structure} $\calz_v=\calz^{V(\gamma(v))}$.
%Considering the partial knowledge model under the existence of a general adversary, 
% $\calz_v = \{ A \cap V(\gamma(v)) \mid A \in \calz \}$(\emph{local adversary structure}).
We denote an instance of the problem by the tuple $I=(G,\calz,\gamma,S,R)$, whereas in the \textsl{ad hoc case} we will omit $\gamma$ since it's fixed. 
%We next define some useful protocol properties. 
%%\begin{definition}[Resilient algorithm for instance $I$]
%%An algorithm which achieves RMT in an instance $\cali$ for any corruption set $Z\in \calz$ and any behavior of $Z$  is called \emph{resilient} for $\cali$.
%%\end{definition}
%%
%%\begin{definition}[Safe / $(\calz, \gamma)$-Safe algorithms]
%%An algorithm which never causes an honest node to decide on (output) an incorrect value, for any instance $\cali$, is called \emph{safe}.\\
%%An algorithm which never causes an honest node to decide on an incorrect value under for any $t$-local corruption set and any behavior of it, for any graph-dealer pair $(G,S)$,  is called \emph{$t$-locally safe}. 
%%\end{definition}
%We say that an RMT protocol is \emph{resilient} for an instance $I$ if it achieves RMT on instance $I$ for any possible corruption set and any admissible behavior of the corrupted players.
 We will say that an RMT protocol is \emph{safe} if it never causes the receiver $R$ to decide on an incorrect value in any instance. The importance of the safety property is pointed out in~\cite{PP05},   
where it is regarded as a basic requirement of a Broadcast algorithm. 
%In RMT, the property guarantees that the receiver will not decide on an incorrect value if she receives any piece of insufficient information. 

%; similarly, in the case of RMT it guarantees that if the receiver does not have sufficient information to decide on the sender's value, she  won't eventually decide on an incorrect value or accept false data.

%
% Given a set of nodes $V$  $(\calz_v)_{v\in V}$,  we say that an RMT protocol is \emph{locally safe} if does not cause a player $v$ to decide on an incorrect value, in every instance that $\calz$ conforms to the local adversary structure $\calz_v$.

%even under an adversary structure which conforms to $\calz_v$.

%
%
%\begin{definition}[Uniqueness of algorithm]
%Let $\cala$ be a family of algorithms. An algorithm $A$ is unique (for RMT) among algorithms in $\cala$ if the existence of an algorithm of family $\cala$ which achieves RMT in an  instance  $I$ implies that $A$ also achieves RMT in $I$.
%\end{definition}
%
%A unique algorithm $A$ among $\cala$, naturally defines the class of instances  in which the problem is solvable by $\cala$-algorithms, namely the ones that $A$ achieves RMT in. 

\section{RMT in the ad hoc model}
In this section we present a tight condition for RMT in the \textsl{ad hoc} model. The proofs are variations of the respective theorems in~\cite{DISTPPS17} for the Broadcast problem. 
The following notion proves crucial for the exact characterization of instances where RMT is possible.\vspace{3pt}

\noindent(\emph{$\calz$-partial pair cut})\hspace{5pt} 
Let $C$ be a cut of $G$ partitioning $V\setminus C$ into sets $A,B\neq \emptyset$ s.t.\ $S\in A$ and $R\in B$.
$C$ is a \emph{$\calz$-pp cut} if there exists a partition $C=C_1\cup C_2$ with $C_1\in \calz$
and $\forall u\in B, \ \caln(u)\cap C_2\in \calz_u$.

Observe that $\calz$-pp cut, constitutes a one-sided variation of the well known $\calq^{(2)}$ separator notion used in~\cite{KGSR02} in the context of full topology knowledge networks. Its existence in a network yields impossibility of safe RMT as shown below.\vspace{-4pt}

\begin{theorem}[RMT Impossibility]\label{thrm:zCPAnec}
 Given an RMT instance $(G,\calz,S,R)$, if an $RMT \ \calz$-pp cut exists on $G$ then no safe RMT algorithm exists for $(G,\calz,S,R)$.
\end{theorem}\vspace{-4pt}

Checking the above condition has been proven in~\cite{DISTPPS17}  to be $\mathrm{NP}$-hard but interestingly, the impossibility is matched by a variation of the simple Broadcast algorithm $\calz$-CPA (\emph{Certified Propagation Algorithm}). The algorithm was introduced in the same work and in turn, can be seen as a generalization of CPA Broadcast algorithm, proposed in~\cite{Koo04}. In $\calz$-CPA, initially,  the sender sends its value $x_S$ to all its neighbors and terminates. Subsequently, each player $v$ decides on a value through a \emph{decision rule} and propagates its decision to all its neighbors if $v\neq R$, or outputs its decision if $v= R$. The core of $\calz$-CPA consists of the following decision rules:\vspace{4pt}

\noindent\textbf{$\calz$-CPA decision rules}: If $v\in \caln(S)$ then upon reception of $x_S$ from the sender, decide on $x_S$.  Else, if $v\notin \caln(S)$ then  upon receiving the same  value 
$x$ from all neighbors in a set $N\subseteq \caln(v)$ such that\ $N\notin \calz_v$,  decide on value $x$. \vspace{3pt}

$\calz$-CPA proves to be tight for RMT by a variation of the theorem presented in~\cite{DISTPPS17} as shown below. This practically means that given an \textsl{ad hoc} RMT instance $\cali$, the best one can do is to execute $\calz$-CPA; if RMT is not achieved then no safe algorithm can achieve RMT in $\cali$. \vspace{-4pt}

 \begin{theorem}[Feasibility of RMT]\label{thrm:zCPAsuf}
    Given an RMT instance $(G,\calz,S,R)$, if no $RMT \ \calz$-pp cut exists on $G$,  then $\calz$-CPA achieves RMT in $(G,\calz,S,R)$.
 \end{theorem} \vspace{-4pt}
 
 The minimal communication that $\calz$-CPA utilizes is noteworthy. For instance, Theorems~\ref{thrm:zCPAnec},\ref{thrm:zCPAsuf} imply that in the \textsl{ad hoc} model, local knowledge exchange between players cannot benefit the feasibility of the problem. Considering partial knowledge in its most general form, presents a new challenge in the study; namely, it becomes clear that tight results can be obtained only if the
 players exchange and appropriately combine their knowledge regarding the topology and the adversary.

\section{Joining partial knowledge}
Combining topological knowledge exchanged by two players $v,w$ is trivial, since their joint knowledge $\gamma(v,w)$ can easily be computed. However, their joint knowledge over the adversary structure is not obvious to define.  In this section we address the issue and define a \emph{joint operation} over local adversary structures\footnote{The operation takes into account potentially different adversarial structures, so that it is well defined even if a corrupted player provides a different structure than the real one to
some honest player.}. As is proved in the following, this operation allows the combination of local knowledge in an optimal way. Details of this section's  results can be found in~\cite{PPS17}. \vspace{3pt}

%
%In this setting each player $v$ only has knowledge of the topology of a certain connected subgraph $G_v$ of $G$
%which includes $v$. Namely if we consider the family $\calg$ of connected subgraphs of $G$
%we use the \emph{view function } $\gamma: V\rightarrow \calg$, where $\gamma(v)$ represents the subgraph 
%over which player $v$ has knowledge of the topology.
%We extend the domain of $\gamma$ by allowing as input a set $S \subseteq G$. The output will correspond to the joint view of nodes in $S$.
% each player $v$ knows the possible corruption sets in his view $\calz_u = \{ z \cap V(\gamma(u)) \mid z \in \calz \}$.

%Considering two players who have partial knowledge of the adversary, we define an operation to calculate their joint knowledge about the adversary. 
%For an adversary structure $\cale$ and a node set $A$ let $\cale^A = \{ Z \cap A \mid  Z \in \cale \}$. The joint adversary structure from two restricted adversary structures can be obtained through the  $\oplus$ operator\footnote{ We define the operation on two possibly different structures $\cale, \calf$ so that the operation is well defined even if a corrupted player falsely reports a different structure than the real one.}.

%We are interested on properties of this operation on different structures because the adversary can pretend having a different structure than the real one. 
%Let $\cale,\calf,\calh$ be adversary structures and $A,B,C$ be sets of nodes.

\noindent(\emph{Joint operation $\oplus$})\hspace{5pt}
Let $\mathbb{T}^A=2^{2^A}$ denote the space of adversary structures on a set of nodes $A$. For any node sets $A,B$ and adversary structures $\cale, \calf$, the  operation  $\oplus:\mathbb{T}^A \times \mathbb{T}^B \rightarrow \mathbb{T}^{(A\cup B)}$, is defined as follows:
$$ \cale^A \oplus \calf^B  = \{Z_1 \cup Z_2| (Z_1 \in \cale^A) \wedge (Z_2 \in \calf^B) \wedge (Z_1 \cap B = Z_2 \cap A) \} 
$$

Informally, the $\cale^A \oplus \calf^B$ operation  unites possible corruption sets from $\cale^A$ and $\calf^B$ that `agree' on $A\cap B$. The $\oplus$ operation is commutative, associative and idempotent, which means that it imposes a \emph{semi-lattice structure} on the space of all possible partial adversary information. The semi-lattice structure allows for the definition of a partial order relation $\succcurlyeq$ on this space which guarantees the existence of a supremum for every finite subset under $\succcurlyeq$. Utilizing the latter, we can prove the following property of the $\oplus$ operation which is important for our study.\vspace{-4pt}
%
%\begin{lemma}\label{[lemma:locindist]}
%For any adversary structures $\cale,\calf$, node sets $A,B$
%and $\calh = \cale^A \oplus \calf^B$, it holds  that  $\calh^A = \cale^A$ and $\calh^B = \calf^B$.
%\end{lemma}
%
%\begin{theorem}\label{[theorem:maxinfo]}
%For any adversary structures $\cale,\calf$, node sets $A,B$
%and $\calh = \cale^A \oplus \calf^B$, it holds that $\forall \calh' \in \mathbb{T}^{A\cup B}$ :
%if $\calh'^A = \cale^A$ and $\calh'^B = \calf^B$ then $\calh' \subseteq \calh$.
%\end{theorem}
%%
%\begin{proof}
%Suppose that there existed some $\calh'$ s.t.\ $\exists Z\in \calh': Z\not \in \calh$. For $Z$ we have $Z_1 = Z\cap A \in \cale^A$ and $Z_2= Z\cap B \in \calf^B$. Also $Z_1 \cap B = Z \cap A \cap B =Z_2\cap A$. But then, definition~\ref{oplusdefinition} implies $Z \in \calh$, a contradiction. 
%\end{proof} 
%
\begin{theorem}\label{cor:adversaryineq}
For any adversary structure $\calz$ and node sets $A,B$:
$\calz^{(A \cup B)} \subseteq \calz^A \oplus \calz^B  $.
\end{theorem}\vspace{-4pt}
What Theorem~\ref{cor:adversaryineq} shows is that the $\oplus$ operation gives the maximal (w.r.t inclusion) possible adversary structure  that is indistinguishable 
by two agents that know $\calz^A$ and $\calz^B$ respectively, i.e., it coincides with their knowledge of the adversary structures on sets $A$ and $B$ respectively.
%Recall that  $\calz_u = \calz^{V(\gamma(u))}$. We will prefer to use $\calz_u$ to denote the local adversary structure of player $u$ and  $\calz^{V(\gamma(u))}$ to denote the corresponding restriction of the adversary structure. 
We can now define the combined knowledge of a set of nodes $B$ about the adversary structure $\calz$ given a view function $\gamma$:\vspace{-3pt}
$$
\calz_B = \bigoplus\limits_{v \in B} \calz^{V(\gamma(v))}
$$\vspace{-7pt}

Note that $\calz_B$ exactly captures the maximal adversary structure possible, restricted in $\gamma(B)$, conforming to the initial knowledge of players in $B$. Also notice that using Theorem~\ref{cor:adversaryineq} we get  $\calz^{V(\gamma(B))} \subseteq \calz_B$. This means that, what nodes in $B$ conceive as the worst case adversary structure indistinguishable to them, 
%relative to their initial knowledge
  always contains the actual adversary structure in their scenario.

Impossibility of RMT in this case is based on the following separator notion  which is analogous to that of $\calz$-pp cut but also involves the joint knowledge of players.  \vspace{3pt}
 
  \noindent(\emph{RMT-cut})\hspace{5pt}  
Let $(G,\calz,\gamma,S,R)$ be an RMT instance and $C=C_1 \cup C_2$ be a cut in $G$, partitioning $V \setminus C$ in two sets $A,B \neq \emptyset$ where $S \in A$ and $R \in B$. Then $C$ is a  \emph{RMT-cut}  iff  $C_1 \in \calz$ and  $C_2\cap V(\gamma(B)) \in \calz_B$.\vspace{3pt}

\noindent
Then the impossibility result can be stated as follows. \vspace{-4pt}
%The proof combines ideas from~\cite{PP05,PPS14} with the $\oplus$ operation and is presented  in~\cite{PPS15}.

%
\begin{theorem}[Necessity]
\label{thm:necessity}
Let $(G,\calz,\gamma,D,R)$ be an  RMT instance.
If there exists a RMT-cut in $G$ then  no safe and resilient RMT algorithm exists for $(G,\calz,\gamma,D,R)$.
\end{theorem}\vspace{-4pt}

\noindent\textbf{Conclusions.} As in the \textsl{ad hoc} case, algorithm \emph{RMT-PKA} proposed in~\cite{PPS17}, is proven tight. This proves the optimality of the local information exchange procedure. On the negative, RMT-PKA is of exponential bit complexity and the reduction of the communication cost is an open problem. Also an interesting research direction would be to study the exact threshold which renders local information exchange useful.

%\input{RMTreduction}

%\section{Open Questions}
%%\begin{itemize}
%%\item 
%The RMT-PKA protocol employs  topology information exchange between players. 
%Although topology discovery was not our motive, techniques used here (e.g.\ the $\oplus$ operation) may be applicable to that problem under a Byzantine adversary (e.g.~\cite{NT09}). A comparison with the techniques used in this field might give further insight on how to efficiently extract information from maliciously crafted topological data.
%
%We have shown that RMT-PKA protocol is unique for the partial knowledge model; this only addresses the feasibility issue. A natural question is whether and when we can devise a unique and also efficient algorithm for this setting. The techniques used so far to reduce the communication complexity (e.g.~\cite{KGSR02}) do not seem to be directly applicable to this model. So, exploring this direction further is particularly meaningful. 
%
%It would also be interesting  to argue about uniqueness with respect to efficiency for RMT in the partial knowledge model  by extending our analysis of the \textsl{ad hoc} case. 
%%\end{itemize}

%Finally, it is possible to define a stronger type of poly-time uniqueness: we call a protocol scheme $\cala$ \emph{strongly poly-time unique} for problem $\Pi$ if the existence of any fully-polynomial protocol for a class of instances $\cali$ implies that $\cala$ is also fully polynomial for all instances in $\cali$. We conjecture that $\calz$-CPA is in fact strongly poly-time unique for RMT in the \textsl{ad hoc} model.

\bibliographystyle{abbrv}
\bibliography{Broadcast}

\end{document}